\documentclass[
aps,
prd,
reprint,
onecolumn,
superscriptaddress,
amsmath,
amssymb,
floatfix,
nofootinbib % <--- ESSA LINHA TRAZ AS NOTAS DE VOLTA PARA O RODAPÉ
]{revtex4-2}

\usepackage[utf8]{inputenc}
\usepackage[T1]{fontenc}
\usepackage{microtype}
\usepackage{lmodern}
\usepackage{comment}
\usepackage{placeins} 
\usepackage[english]{babel}
\usepackage{setspace}
\usepackage{amsmath}
\usepackage{amssymb}
\usepackage{amsfonts}
\usepackage{mathtools}
\usepackage{bm}

\usepackage{graphicx}
\usepackage{float}
\usepackage{caption}
\usepackage{subcaption}

\usepackage{epstopdf}

\usepackage{booktabs}
\usepackage{array}
\usepackage{multirow}

\allowdisplaybreaks

\usepackage{xcolor}

\usepackage{physics}

\usepackage{siunitx}

\usepackage[
colorlinks=true,
linkcolor=blue,
citecolor=blue,
urlcolor=blue
]{hyperref}

\begin{document}

% =========================================================
%  TÍTULO
% =========================================================

\title{Stationary Rotating Geometries Associated with Nonsingular Pulsating Collapse}

% =========================================================
%  AUTORES
% =========================================================

\author{A. D. Souza}
\email{almir.souza@professor.pb.gov.br}
\affiliation{
	Department of Physics,
	Federal University of Campina Grande,
	Brazil
}

\author{F. A. Brito}
\email{fabrito@df.ufcg.edu.br}
\affiliation{
	Department of Physics,
	Federal University of Campina Grande,
	Brazil
}

\author{J. Spinelly}
\email{jeanspinelly@servidor.uepb.edu.br}
\affiliation{
	Department of Physics,
	State University of Campina Grande,
	Brazil
}

% =========================================================
%  DATA
% =========================================================

\date{\today}

% =========================================================
%  RESUMO
% =========================================================

\begin{abstract}
	
	Since the first exact solutions of General Relativity (GR) were obtained, it became clear that the theory predicts a new class of compact objects: black holes. The Hawking--Penrose singularity theorems show that, under appropriate assumptions, gravitational collapse in GR leads to singular behavior. Motivated by the expectation that such regimes require physics beyond classical GR, effective approaches inspired by Loop Quantum Cosmology (LQC) and braneworld scenarios introduce high-density corrections. In this work we consider the static exterior geometry introduced by Gao, Lu, Shen, and Faraoni, associated with a comoving interior model in which the singularity is avoided through a sequence of collapse and bounce phases. Using the Newman--Janis algorithm in the Azreg-Aïnou formulation, we construct a stationary rotating exterior extension of this geometry. The resulting line element describes a stationary axisymmetric rotating spacetime whose horizon geometry, angular velocity, surface gravity, and Hawking temperature recover the Kerr and Schwarzschild limits when the quantum/braneworld correction becomes negligible ($l\to0$). We further analyze the thermal response at fixed rotational state-space parameter, identify Davies-type singular points, and formulate an extended thermodynamic state-space relation.

\end{abstract}
	
\maketitle

% =========================================================
%  						INTRODUÇÃO
% =========================================================

\section{Introduction}

General Relativity (GR) describes gravitation as a manifestation of spacetime curvature generated by matter and energy \cite{CARMELI, INVERNO, LIEBSCHER}. The geometry therefore depends on the distribution of matter-energy and determines the corresponding gravitational trajectories and observables \cite{CARMELI, MISNER}. Since the formulation of GR in 1915, exact and approximate solutions of the Einstein field equations have played a central role in understanding both gravitational dynamics and the evolution of the Universe.

One of the earliest and most important examples is the Schwarzschild solution, obtained in 1916. It describes the exterior geometry generated by a static, spherically symmetric source and constitutes the prototype of a black hole spacetime. Other landmark solutions include the Reissner-Nordström, Kerr, and Kerr-Newman metrics \cite{MISNER}. In astrophysical contexts, rotation is an indispensable feature, as most stellar-mass and supermassive compact objects possess non-zero angular momentum. Consequently, stationary axially symmetric spacetimes provide the proper theoretical background for testing gravity in strong-field regimes.

Although black hole solutions arise naturally in GR and are strongly supported by observations such as direct gravitational-wave detections by the LIGO-Virgo-KAGRA Collaboration \cite{Abbott_2016} and event-horizon-scale imaging by the Event Horizon Telescope (EHT) \cite{Akiyama_2019,Akiyama_2022}, the occurrence of intrinsic curvature singularities indicates the breakdown of the classical description in sufficiently high-curvature regimes \cite{SUSSKIND, Penrose_1965}. According to the classical Hawking-Penrose singularity theorems, generic gravitational collapse under reasonable energy conditions inevitably leads to geodesically incomplete spacetimes. 

To overcome this issue, regular (non-singular) black holes and bounce scenarios have been extensively developed in the literature \cite{Bardeen_1968,Simpson_2019,Lemos_2011,Lan_2023,Carballo_Rubio_2018,Franzin_2021,Frolov_2016,Bambi_2013,Malafarina_2017}. These effective models are typically inspired by candidate theories of quantum gravity, such as Loop Quantum Gravity (LQG) \cite{Rovelli_2014,Gambini_2013} and Loop Quantum Cosmology (LQC) \cite{Ashtekar_2011,Agullo_2016}, as well as braneworld gravity scenarios \cite{Maartens_2010}. In these frameworks, classical singularities are circumvented via quantum-geometry effects or higher-dimensional modifications, which introduce repulsive quantum-pressure terms at Planckian scales.

In this context, Gao \textit{et al.} (2018) \cite{CHANGJUN} revisited the classic Oppenheimer-Snyder gravitational-collapse model and extended it to incorporate effective matter contents with modifications originating from braneworld scenarios and LQC dynamics. In their construction, the collapse of a fluid sphere avoids the central singularity: as the interior density approaches a critical scale, the gravitational contraction is halted by effective repulsive forces, triggering a smooth bounce followed by an expansion phase. Subsequent recollapse cycles can give rise to a pulsating or oscillating interior dynamics, offering a well-defined regular alternative to classical singular collapse \cite{CHANGJUN}.

While spherical symmetry offers a tractable framework for gravitational collapse, realistic compact objects inevitably spin. Extending non-singular and quantum-corrected collapse models to rotating geometry is therefore essential for physical applications, including the study of accretion-disk physics, shadows, and quasinormal modes \cite{Bambi_2017,Toshmatov_2017,Tsukamoto_2018,Campos2022,Anacleto2021b}. 

The purpose of the present work is to investigate how rotation modifies the exterior geometry associated with the non-singular collapse model proposed by Gao \textit{et al.} \cite{CHANGJUN}. To construct the spinning counterpart, we employ the Azreg-Aïnou prescription \cite{MUSTAPHA,Azreg_Ainou_2014_EPJC}, a non-complexification formulation of the original Newman--Janis algorithm (NJA) \cite{JANIS}. The classic NJA often introduces mathematical ambiguities when complexifying arbitrary non-vacuum or modified metric functions. The Azreg-Aïnou technique bypasses these limitations by enforcing consistency in the fluid source structure without requiring ad hoc complex transformations.

The resulting stationary, axisymmetric exterior spacetime will be referred to as the Rotating Exterior Metric (REM). We examine the horizon structure, ergoregions, and geometric properties of REM as a function of both the spin parameter and the quantum/braneworld correction scale. As expected, in the limit where the characteristic effective parameter vanishes, the standard classical rotating black hole geometries are smoothly recovered. We use units in which $G=c=\hbar=k_B=1$ and the metric signature $(-,+,+,+)$.

% =========================================================
%  			Pulsating-collapse model with rotation
% =========================================================

\section{Stationary Rotating Exterior geometry  Associated with the pulsating-collapse metric}

In this section we apply the modified Newman--Janis construction to obtain a rotating line element. We start from the static exterior metric of Gao \textit{et al.} (2018) \cite{CHANGJUN}:
\begin{eqnarray}
	ds^2= -F(x) d\tau^2 + F(x)^{-1} dx^2 + x^2d\Omega^2, 			\label{I-1}
\end{eqnarray}
where $d\Omega^2=d\theta^2+\sin^2(\theta)d\psi^2$, with
\begin{eqnarray}
	F(x)=1-\frac{2M_{f}}{x^{f}}\left(1-\frac{l^{f+2}}{x^{f+2}}\right),		\label{I-2}
\end{eqnarray}
\begin{eqnarray}
	M_{f} \equiv \frac{4\pi}{3}\rho_{{}_f}r_{0}^{f+2} 	\label{I-2.1}
\end{eqnarray}
and
\begin{eqnarray}
	l^{f+2} \equiv \frac{\rho_{{}_f}r_{0}^{f+2}}{\rho_{cr.}}\ .		\label{I-2.2}
\end{eqnarray}
Here $r_0$ is the proper radius of the fluid sphere and $f$ takes the values $f=1,2,4$. The quantities $\rho_1$, $\rho_2$, and $\rho_4$ denote the densities associated, respectively, with dust, radiation, and stiff matter, while $\rho_{\rm cr.}$ is a constant of the order of the Planck energy density.

According to the Newman--Janis algorithm \cite{JANIS, NEWMAN}, we first perform the transformation
\begin{eqnarray}
	d\tau=du+\frac{dx}{F(x)},			\label{I-3}
\end{eqnarray}
which introduces a null time coordinate and yields
\begin{eqnarray}
	ds^2=-F(x)du^2 - 2dudx + x^2d\Omega^2. 		\label{I-4}
\end{eqnarray}
By using the coordinate ordering $x^\mu=(u,x,\theta,\psi)$, the contravariant metric $g^{\mu\nu}$ takes the form
\begin{eqnarray}
	g^{\mu\nu} = \begin{pmatrix}  0 & -1 & 0 & 0 \\  -1 & F(x) & 0 & 0 \\  0 & 0 & \frac{1}{x^2} & 0 \\  0 & 0 & 0 & \frac{1}{x^2 \sin^2\theta}  \end{pmatrix}.
	\label{I-4.1}
\end{eqnarray}
In terms of the null tetrad $(l^\mu,n^\mu,m^\mu,\bar m^\mu)$, the inverse metric can be written as
\begin{eqnarray}
	g^{\mu\nu} = -l^\mu n^\nu - l^\nu n^\mu + m^\mu \bar{m}^\nu + m^\nu \bar{m}^\mu
	\label{I-5}
\end{eqnarray}
with the following null tetrad vectors for the pulsating-collapse spacetime:
\begin{eqnarray}
	l^{\mu}=\delta^{\mu}_{x} \ ,
	\label{I-5.1}
\end{eqnarray}

\begin{eqnarray}
	n^{\mu}= \delta^{\mu}_{u}-\frac{1}{2}F(x)\delta^{\mu}_{x} \ ,
	\label{I-5.2}
\end{eqnarray}
\begin{eqnarray}
	m^{\mu} = \frac{1}{x\sqrt{2}}\left(\delta^{\mu}_{\theta}+\frac{i}{\sin{\theta}}\delta^{\mu}_{\psi}\right)
	\label{I-5.3}
\end{eqnarray}
and
\begin{eqnarray}
	\bar{m}^{\mu} = \frac{1}{x\sqrt{2}}\left(\delta^{\mu}_{\theta}-\frac{i}{\sin{\theta}}\delta^{\mu}_{\psi}\right).
	\label{I-5.4}
\end{eqnarray}
Following the Newman--Janis prescription, we introduce the complex coordinate transformation
\begin{eqnarray}
	x \rightarrow \tilde{x}=x+i a \cos(\theta), \quad	u \rightarrow \tilde{u}=u+i a \cos(\theta), \quad \tilde\theta = \theta, \quad \tilde\psi = \psi,
	\label{I-6}
\end{eqnarray}
where $a$ is the rotation parameter, with dimensions of length.
In the dust sector ($f=1$), for which $M_1$ has the standard
asymptotic mass interpretation, one recovers the usual identification
$J=aM_1$. For $f=2$ and $f=4$, $M_f$ is instead regarded as an
effective parameter of the radial matter profile, and no standard
asymptotic identification $J=aM_f$ is assumed.

%%%%%%%%%%%%

In the Azreg-Aïnou extension \cite{MUSTAPHA}, rather than postulating an arbitrary complexification of $F(x)$, one introduces generic real functions of $(x,\theta,a)$, denoted by $A(x,\theta)$ and $\Psi(x,\theta)$:
\begin{eqnarray}
	x^2 \to \Psi(x,\theta) = \Sigma = x^2 + a^2 \cos^2\theta,  
	\label{I-8}
\end{eqnarray}
\begin{eqnarray}
	F(x) \to A(x, \theta) = \frac{F(x) x^2 + a^2 \cos^2\theta}{\Sigma}.
	\label{I-8.1}
\end{eqnarray}
The transformed null tetrads acquire an angular dependence, a dependence on the rotation parameter $a$ and can be written as
\begin{eqnarray}
	\tilde{l}^{\mu}=\delta^{\mu}_{x},
	\label{I-9}
\end{eqnarray}
\begin{eqnarray}
	\tilde{n}^{\mu}= \delta^{\mu}_{u} - \frac{1}{2} A(x, \theta) \delta^{\mu}_{x},
	\label{I-9.1}
\end{eqnarray}
\begin{eqnarray}
	\tilde{m}^{\mu} = \frac{1}{\sqrt{2\Sigma}}\left(\delta^{\mu}_{\theta}+ia \sin\theta (\delta^{\mu}_{u}-\delta^{\mu}_{x})+ \frac{i}{\sin{\theta}}\delta^{\mu}_{\psi}\right)
	\label{I-9.2}
\end{eqnarray}
and
\begin{eqnarray}
	\tilde{\bar{m}}^{\mu} = \frac{1}{\sqrt{2\Sigma}}\left(\delta^{\mu}_{\theta}-ia \sin\theta (\delta^{\mu}_{u}-\delta^{\mu}_{x})- \frac{i}{\sin{\theta}}\delta^{\mu}_{\psi}\right).
	\label{I-9.3}
\end{eqnarray}
Computing the metric components from Eq.~(\ref{I-5}) and taking the inverse matrix gives
\begin{eqnarray}
	g_{\mu\nu} = \begin{pmatrix} -A(x,\theta) & -1 & 0 & a \sin^2\theta \left(1 - A(x,\theta)\right) \\ -1 & 0 & 0 & a \sin^2\theta \\ 0 & 0 & \Sigma & 0 \\ a \sin^2\theta \left(1 - A(x,\theta)\right) & a \sin^2\theta & 0 & \sin^2\theta \left[\Sigma + a^2 \sin^2\theta \left(2 - A(x,\theta)\right)\right] \end{pmatrix},
	\label{I-10}
\end{eqnarray}
which leads to the following line element in Eddington--Finkelstein coordinates:
\begin{eqnarray}
	ds^2 = -A(x,\theta) du^2 - 2 du dx + 2 a \sin^2\theta (1 - A(x,\theta)) du d\psi + \nonumber\\
	+ 2 a \sin^2\theta dx d\psi + \Sigma d\theta^2 + \sin^2\theta \left[\Sigma + a^2 \sin^2\theta (2 - A(x,\theta))\right] d\psi^2.
	\label{I-11}
\end{eqnarray}

To express the metric in canonical Boyer--Lindquist coordinates $(t,x,\theta,\phi)$, we use $du=dt-\lambda(x)dx$ and $d\psi=d\phi-\gamma(x)dx$. Requiring the coefficients of $dt\,dx$ and $d\phi\,dx$ to vanish identically yields a system for $\lambda(x)$ and $\gamma(x)$, whose solution is
\begin{eqnarray}
	\lambda(x) = \frac{ x^2 + a^2}{x^2 F(x) + a^2 } \quad\quad {\rm and} \quad\quad \gamma(x) = \frac{a}{x^2 F(x) + a^2}.
	\label{I-12}
\end{eqnarray}

The Azreg-Aïnou construction requires the $dt\,dx$ and $d\phi\,dx$ cross terms to vanish identically. Substituting $A(x,\theta)=\frac{x^2F(x)+a^2\cos^2\theta}{\Sigma}=\frac{\Delta(x)-a^2\sin^2\theta}{\Sigma}$ gives the exact cancellation of these unwanted cross terms. The stationary rotating metric in Boyer--Lindquist coordinates is therefore
\begin{eqnarray}
	ds^2 = -\left(1 - \frac{2 M(x) x}{\Sigma}\right) dt^2 - \frac{4 a M(x) x \sin^2\theta}{\Sigma} dt \, d\phi + \frac{\Sigma}{\Delta(x)} dx^2 +
	\nonumber\\
	+\Sigma \, d\theta^2 + \left(x^2 + a^2 + \frac{2 a^2 M(x) x \sin^2\theta}{\Sigma}\right) \sin^2\theta \, d\phi^2,
	\label{I-13}
\end{eqnarray}
where the structural functions of the spacetime are:
\begin{itemize}
	\item Geometrical function ($\Sigma$):
\end{itemize}
\begin{eqnarray}
	\Sigma = x^2 + a^2 \cos^2\theta.
	\label{I-14}
\end{eqnarray}
\begin{itemize}
	\item Effective mass function $M(x)$:
\end{itemize}
\begin{eqnarray}
	M(x) = \frac{M_f}{x^{f-1}} \left(1 - \frac{l^{f+2}}{x^{f+2}}\right).
	\label{I-14.1}
\end{eqnarray}
\begin{itemize}
	\item Radial function ($\Delta$):
\end{itemize}
\begin{eqnarray}
	\Delta(x) = x^2 + a^2 - 2 M(x) x = x^2 + a^2 - \frac{2M_f}{x^{f-2}} \left(1 - \frac{l^{f+2}}{x^{f+2}}\right).
	\label{I-14.2}
\end{eqnarray}

Equation~(\ref{I-13}) contains a single off-diagonal term, $g_{\phi t}=g_{t\phi}$, which encodes frame dragging and therefore the rotational character of the spacetime. The parameter $l$ carries the effective braneworld/LQC correction. Thus, the line element describes a stationary rotating exterior spacetime associated with a matter distribution whose interior dynamics is governed by the pulsating Gao \textit{et al.} (2018) \cite{CHANGJUN} model.

For $a=0$, $f=1$, and $l\ll x$, the Schwarzschild limit is recovered. For $a\neq0$, $f=1$, and $l\ll x$, the metric approaches Kerr. When $a\neq0$, $f=1$, and the scale $l$ is not negligible compared with $x$, the geometry retains a Kerr-like stationary structure while receiving finite-$l$ corrections. The sectors $f=2$ and $f=4$ have their own radial behavior because of both the effective high-density correction and the different matter equations of state.

It is worth emphasizing that, within classical GR, the standard collapse model does not provide a nonsingular pulsating black hole interior: in the absence of a limiting density, the contraction can proceed toward the singular regime, as in the Oppenheimer--Snyder model \cite{OPPENHEIMER}.

In the effective braneworld/LQC description, the high-density correction can instead bound the collapse. In the regime in which $l$ is much smaller than the horizon scale, the exterior approaches the corresponding classical rotating geometry.

%%%%%%%%%%    Infinite-redshift surfaces and horizons       %%%%%%%%%%%%

\subsection{Infinite-redshift Surfaces and Horizons}

We now analyze the characteristic hypersurfaces of the Rotating Exterior Metric (REM), beginning with the infinite-redshift surface. It is obtained from \cite{INVERNO, 2CARROLL}:
\begin{eqnarray*}
	g_{tt}=0 \ \Rightarrow \ \ 1 - \frac{2 M(x) x}{\Sigma}=0 \ \Rightarrow \ \ 
\end{eqnarray*}
\begin{eqnarray}
	x^2 - 2 M(x) x + a^2 \cos^2\theta = 0,
	\label{II-1}
\end{eqnarray}
where $M(x)$ is given by Eq.~(\ref{I-14.1}) and contains the effective braneworld/LQC correction.

Because the term $a^2\cos^2\theta$ depends on latitude, the outer stationary-limit surface $x_E$ depends on $\theta$. Since $\cos^2\theta$ ranges from $0$ to $1$, two limiting regions are especially useful:
\begin{itemize}
	\item At the poles ($\theta=0$ and $\theta=\pi$):
\end{itemize}
Since $\cos^2(0)=\cos^2(\pi)=1$, the equation becomes $x^2-2M(x)x+a^2=0$, identical to the horizon equation. Thus,  the stationary-limit surface meets the outer horizon at the poles, $x_E=x_+$.
\begin{itemize}
	\item At the equator ($\theta=\pi/2$):
\end{itemize}
Since $\cos(\pi/2)=0$, the $a^2$ term vanishes and one obtains $x=2M(x)$. The stationary-limit radius is then maximal and lies outside the outer horizon, $x_E>x_+$.

The region between the stationary-limit surface and the outer horizon is the ergoregion. As in Kerr, an observer inside this region cannot remain static with respect to infinity and is forced to co-rotate with the geometry. The existence of the ergoregion also permits, in principle, rotational-energy extraction through the Penrose process.

\begin{figure}[htbp]
	\caption{Two-dimensional cross section: effective braneworld/LQC horizons compared with the classical cases.}
	\centering
	\includegraphics[width=0.9\textwidth, height=12cm, keepaspectratio]{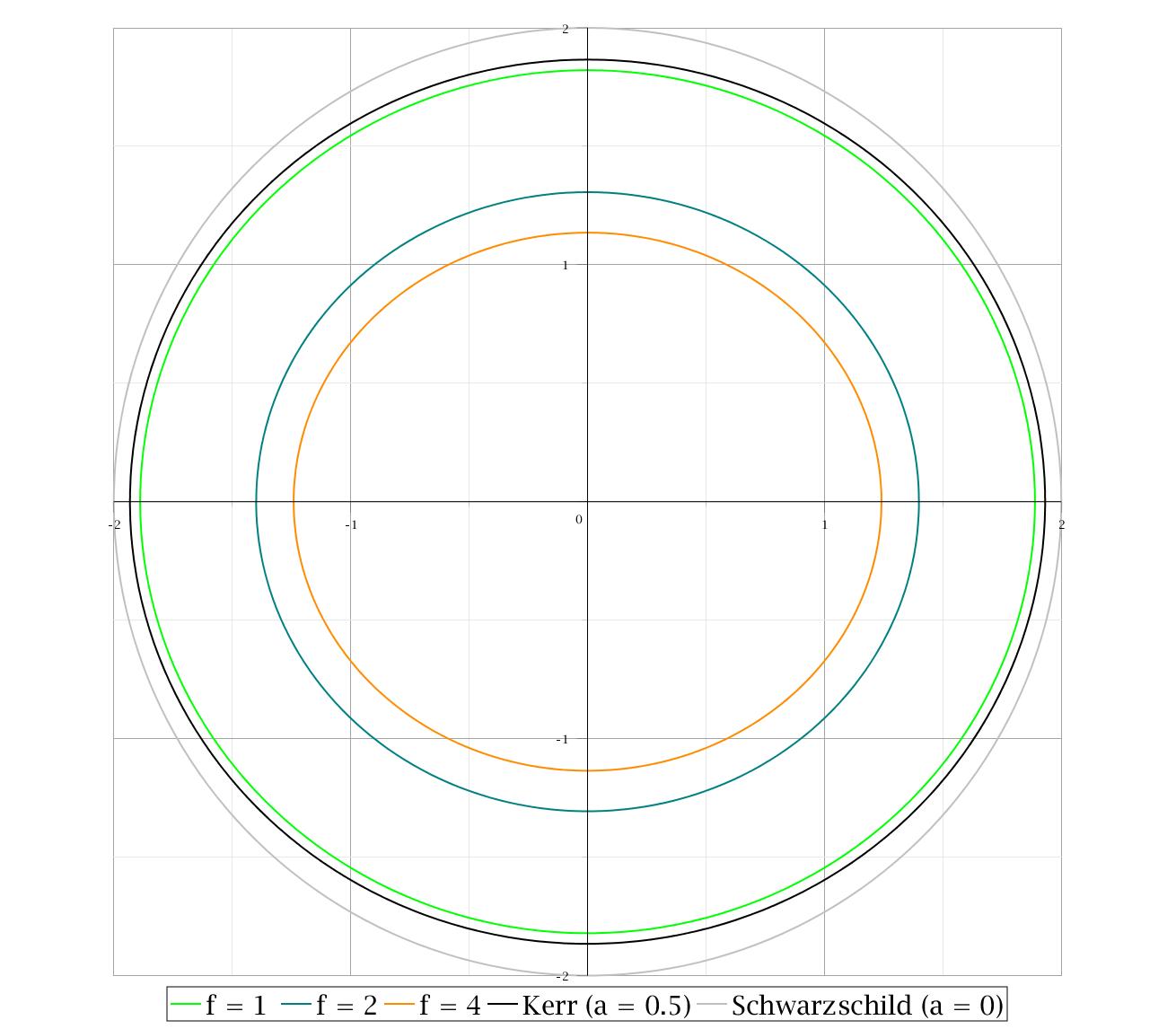}
	\label{horizontes-1}
\end{figure}

The outer horizon is obtained by imposing $\Delta(x_+)=0$ in Eq.~(\ref{I-14.2}) \cite{CARMELI, INVERNO, 2CARROLL, WALD}:
\begin{eqnarray}
	\Delta(x_+) = 0 \quad \Rightarrow \quad x_+^2 + a^2 - \frac{2M_f}{x_+^{f-2}}\left(1 - \frac{l^{f+2}}{x_+^{f+2}}\right) = 0,
	\label{II-2}
\end{eqnarray}
where $x_+$ denotes the radius of the outer horizon.
Figure~\ref{horizontes-1} compares the horizon structure of the modified model with the classical Kerr and Schwarzschild cases. We use $M=1.0$, $a=0.5$, and $l=0.5$. Since Eq.~(\ref{II-2}) cannot be solved for $x_+$ in a single closed form valid for all $f$, the curves are obtained parametrically.
\FloatBarrier
For the dust sector ($f=1$), the outer horizon is shifted relative to Kerr. In the parameter set displayed in Fig.~\ref{horizontes-1}, the $f=2$ and $f=4$ horizons occur at still smaller radii. This behavior reflects both the negative finite-$l$ contribution to the effective mass profile and the different radial falloffs associated with each value of $f$.

%%%%%%%%%%    velocidade angular do BHPR       %%%%%%%%%%%%

\subsection{Angular Velocity of the REM}

We next analyze the angular velocity associated with the REM. Consider a photon emitted in the $\hat\phi$ direction at radial position $x$. At the emission event its momentum has no $\hat x$ or $\hat\theta$ component, and the null condition is \cite{INVERNO, 2CARROLL}:
\begin{eqnarray}
	ds^2=dx=d\theta=g_{tt}dt^2+g_{t\phi}(dt d\phi+d\phi dt)+g_{\phi\phi}d\phi^2=0.
	\label{III-1}
\end{eqnarray}
Solving this equation gives \cite{2CARROLL}
\begin{eqnarray}
	\frac{d\phi}{dt}=-\frac{g_{t \phi}}{g_{\phi\phi}}\pm\sqrt{\left(\frac{g_{t \phi}}{g_{\phi\phi}}\right)^2-\frac{g_{tt}}{g_{\phi\phi}}}.
	\label{angular.0}
\end{eqnarray}
Evaluating Eq.~(\ref{angular.0}) on the outer horizon,
where $\Delta(x_+)=0$ and hence
$g_{t\phi}^{\,2}-g_{tt}g_{\phi\phi}=0$, the two roots
coincide:
\begin{equation}
 \left.\frac{d\phi}{dt}\right|_{x=x_+}
 =
 -\left.\frac{g_{t\phi}}{g_{\phi\phi}}\right|_{x=x_+}
 \equiv \Omega_H .
 \label{angular.1}
\end{equation}
From Eq.~(\ref{I-13}),
\begin{eqnarray}
	g_{t\phi} = -\frac{2 a M(x) x \sin^2\theta}{\Sigma}
	\label{III-18}
\end{eqnarray}
and
\begin{eqnarray}
	g_{\phi\phi} = \left( x^2 + a^2 + \frac{2 a^2 M(x) x \sin^2\theta}{\Sigma} \right) \sin^2\theta
	\label{III-19}
\end{eqnarray}
a straightforward algebra gives
\begin{eqnarray}
	\Omega_H = \frac{a}{x_+^2 + a^2}.
	\label{III-20}
\end{eqnarray}
Equation~(\ref{III-20}) gives the angular velocity of the outer REM horizon. The $\theta$ dependence cancels exactly on the horizon, so the horizon rotates rigidly. Although the functional form is the same as in Kerr, the braneworld/LQC correction affects $\Omega_H$ implicitly through the modified horizon radius $x_+$ determined by Eq.~(\ref{II-2}).

%%%%%%%%%%%%%%%%%  REM thermodynamics  %%%%%%%%%%%%%%%%%%%%

\subsection{Thermodynamics of the REM}

We now examine the laws of black hole mechanics in the REM. We first recall their standard GR formulation \cite{BARDEEN, WALD}.

\textbf{The zeroth law} states that the surface gravity of a stationary black hole is constant over its event horizon.

\textbf{The first law} relates the variation of the black hole mass $M$ to variations of its area $\mathbf{A}$, angular momentum $J$, and electric charge. We restrict attention to the uncharged case, for which \cite{BARDEEN}:
\begin{eqnarray}
	\delta M=\frac{1}{8\pi}\kappa\delta\textbf{A} +\Omega_{H}\delta J_H.
	\label{lei-1}
\end{eqnarray}
Here $\kappa$ is the surface gravity, $\mathbf{A}$ is the horizon area, $\Omega_H$ is the horizon angular velocity, and $J_H$ is the horizon angular momentum.

\textbf{The second law} is the Hawking's area theorem: under the standard classical assumptions, the area $\mathbf{A}$ of a black hole event horizon cannot decrease.\footnote{In ordinary thermodynamics entropy may be transferred between subsystems as long as the total entropy does not decrease. black hole mechanics is more restrictive because a classical black hole cannot bifurcate \cite{HAWKING}; consequently the area theorem applies to each horizon under its stated assumptions \cite{BARDEEN}.} Quantum effects motivate the generalized second law \cite{BEKENSTEIN,HAWKING}:

{\it\textbf{Generalized second law}}: the sum of the black hole entropy and the ordinary entropy outside the black hole does not decrease \cite{BEKENSTEIN}. Thus \cite{1BEKENSTEIN}:
\begin{eqnarray}
	\delta\left(\mathcal{S}_{ext.}+\mathcal{S}_{H}\right)\geq 0.
	\label{lei-2}
\end{eqnarray}

\textbf{The third law} states that the surface gravity $\kappa$ cannot be reduced to zero by any finite sequence of physical operations \cite{BARDEEN}.

These laws closely parallel ordinary thermodynamics, with the usual identifications \cite{2CARROLL,WALD},
\begin{eqnarray}
	T_H=\frac{\kappa}{2\pi} \quad \hbox{and} \quad  \mathcal{S}_{H}=\frac{\textbf{A}}{4},
	\label{T-S}
\end{eqnarray}
where $T_H$ is the Hawking temperature and $\mathcal{S}_H$ is the Bekenstein--Hawking entropy.

We now examine these relations for the REM, where the finite-$l$ terms encode the effective braneworld/LQC modification.

%%%%%%%%%%%%%%%%%%%%%%%%%%%%%%%%%%%%%%%%	

\quad $\ast$ {\bf\textit{Zeroth Law}}

The zeroth law can be checked directly from the REM geometry. We compute the surface gravity from the metric in Eq.~(\ref{I-13}) using \cite{WALD,POISSON}:
\begin{eqnarray}
	\left(\partial_\mu\Phi\right)_{H}=-2\kappa\left(\chi_\mu\right)_{H},
	\label{g-superficie}
\end{eqnarray}
where $\chi_\mu$ is associated with the horizon-generating Killing vector and $\Phi=g_{\mu\nu}\chi^\mu\chi^\nu$, which vanishes on the horizon.

Since the REM is stationary and axisymmetric, the horizon generator is
\begin{eqnarray*}
	\chi^\mu=K^\mu+\Omega_{H}R^\mu,
\end{eqnarray*}
where $\Omega_H$ is the angular velocity of the outer horizon, $K^\mu=\delta^\mu_t$, and $R^\mu=\delta^\mu_\phi$. Hence
\begin{eqnarray}
	\Phi=g_{tt}+2\Omega_{H}g_{t\phi}+\Omega_{H}^2g_{\phi\phi}
	\label{g-superficie1.1}
\end{eqnarray}
and
\begin{eqnarray}
	\chi_\mu=g_{t\mu}+\Omega_{H}g_{\phi\mu},
	\label{g-superficie1.2}
\end{eqnarray}
which leads to
\begin{eqnarray}
	\kappa = \frac{\Delta'(x_+)}{2(x_+^2 + a^2)}
	\label{III-17}
\end{eqnarray}
with
\begin{eqnarray}
	\Delta'(x) = 2x + \frac{2M_f(f-2)}{x^{f-1}} - \frac{4f M_f l^{f+2}}{x^{2f+1}}
	\label{III-17.1}
\end{eqnarray}
and, on the horizon,
\begin{eqnarray}
	\Delta'(x_+) = 2x_+ + \frac{x_+^2 + a^2}{x_+ \left(x_+^{f+2} - l^{f+2}\right)} \left[ (f-2)x_+^{f+2} - 2f l^{f+2} \right].
	\label{III-17.2}
\end{eqnarray}
Therefore, the REM surface gravity is
\begin{eqnarray}
	\kappa = \frac{2 x_+}{2(x_+^2 + a^2)} +\frac{(f-2)x_+^{f+2} - 2f l^{f+2}}{2 x_+ \left(x_+^{f+2} - l^{f+2}\right)},
	\label{III-21}
\end{eqnarray}
which is independent of $\theta$ and is consequently constant over the horizon, in agreement with the zeroth law.\\

%%%%%%%%%%%%%%%%%%%%%%%%%%%%%%%%%%%%%%%%%%%%%%%%

\quad $\ast$ {\bf\textit{Third Law: Extremal Limit}}

We do not attempt to prove the third law in its full dynamical form. We restrict ourselves to identifying the extremal condition.\footnote{The extremal limit is defined by $\kappa_H=0$, and hence $T_H=0$, which determines an extremal horizon radius $x_+^{\rm ext}$. This identifies the zero-Hawking-temperature boundary of the parameter space. No claim of a dynamical proof of the third law is made.} From Eqs.~(\ref{III-21}) and (\ref{T-S}), the Hawking temperature of the REM is
\begin{eqnarray}
	\mathcal{T}_{{}_{REM}}= \frac{1}{4\pi} \left[ \frac{2x_+}{x_+^2 + a^2} + \frac{(f-2)x_+^{f+2} - 2f l^{f+2}}{x_+ \left(x_+^{f+2} - l^{f+2}\right)} \right].
	\label{temp.modificada}
\end{eqnarray}
For $a=0$ one recovers the corresponding Gao \textit{et al.} (2018) \cite{CHANGJUN} result. The classical Kerr expression follows for $f=1$, $a\neq0$, and $l\to0$, while Schwarzschild is recovered by additionally taking $a=0$.

The zeroes of the Hawking temperature, $\mathcal{T}_{\rm REM}=0$, provide the extremality condition when considered together with the horizon equation $\Delta(x_+)=0$. In the usual rotating-horizon picture, extremality corresponds to the degeneracy of the outer and inner horizon branches and to vanishing surface gravity, $\kappa_H=0$. After setting the temperature to zero and simplifying the denominators, the extremal radius is governed by
\begin{eqnarray}
	f x_+^{f+4} + a^2(f-2) x_+^{f+2} - 2(f+1) l^{f+2} x_+^2 - 2f a^2 l^{f+2} = 0 .
	\label{cond-1}
\end{eqnarray}

This polynomial makes explicit the combined influence of the rotation parameter $a$, the matter sector $f$, and the effective correction scale $l$. For dust ($f=1$), it reduces to
\begin{eqnarray}
	x_+^5 - a^2 x_+^3 - 4 l^3 x_+^2 - 2 a^2 l^3 = 0 .
	\label{cond-2}
\end{eqnarray}

In the limit without the effective correction, $l\to0$, this becomes
\begin{eqnarray}
	x_+^3 (x_+^2 - a^2) = 0 ,
	\label{cond-3}
\end{eqnarray}
whose positive root exactly recovers the classical Kerr extremal radius,
$x_+^{\rm ext}=a=M$. A finite value $l\neq0$ continuously shifts this root and therefore modifies the extremal-horizon radius relative to the GR result.

For radiation ($f=2$) and stiff matter ($f=4$), respectively, the algebraic conditions become
\begin{eqnarray}
	x_+^6 - 3 l^4 x_+^2 - 2 a^2 l^4 = 0
	\label{cond-4}
\end{eqnarray}
and
\begin{eqnarray}
	2 x_+^8 + a^2 x_+^6 - 5 l^6 x_+^2 - 4 a^2 l^6 = 0 .
	\label{cond-5}
\end{eqnarray}

Real positive roots, when compatible with the horizon condition $\Delta(x_+)=0$, identify extremal configurations in the corresponding sectors. Physically, the determination of $x_+^{\rm ext}$ establishes the lower boundary of the corresponding horizon branch and delimits the admissible region $x_+\geq x_+^{\rm ext}$ within the parameter range considered.

In the extremal configuration, the vanishing of the surface gravity implies a vanishing Hawking temperature and therefore suppresses semiclassical thermal Hawking emission. Moreover, since the thermal-response function considered below contains an overall factor of $T_H$, the condition $\mathcal{T}_{\rm REM}=0$ implies that this response vanishes at extremality, provided that the remaining factors are regular. This indicates a vanishing Hawking thermal response at the extremal boundary. We therefore interpret the extremal configuration as the zero-Hawking-temperature lower boundary of the allowed parameter space, rather than as a proven globally stable ground state of minimum energy.

\begin{figure}[htbp]
	\caption{Comparison between the classical and modified Hawking temperatures.}
	\centering
	\includegraphics[width=0.9\textwidth, height=12cm, keepaspectratio]{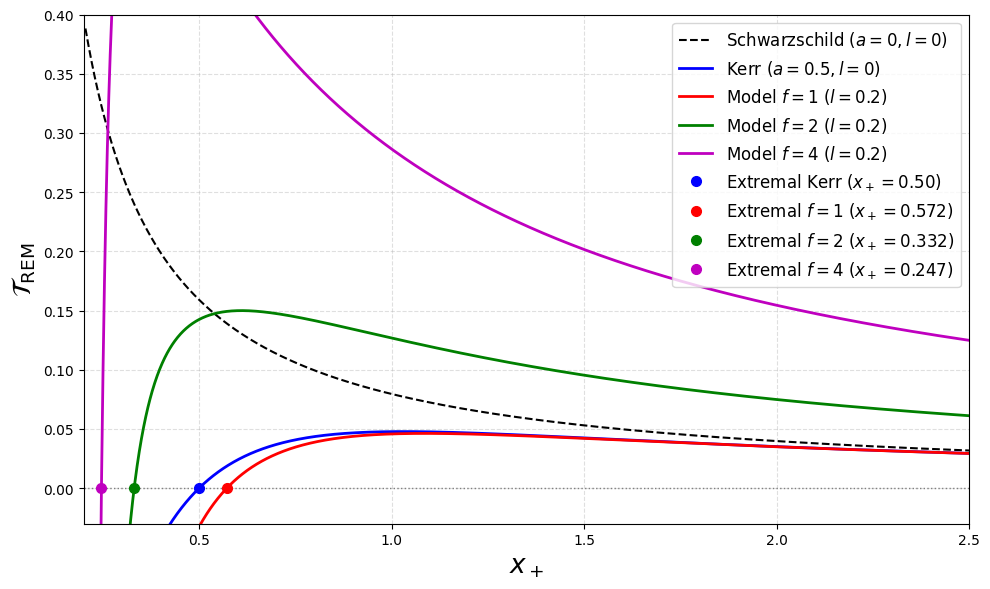}
	\label{extremal-1}
\end{figure}

Figure~\ref{extremal-1} compares the temperatures of the modified model with those of the corresponding classical GR solutions.
\FloatBarrier

A finite correction scale $l\neq0$ appreciably modifies the thermal profile relative to the classical Kerr solution ($l=0$). For the parameters shown, the effective correction reduces the maximum temperature and continuously shifts the characteristic radius at which the thermal response changes. The effect becomes more pronounced when the horizon scale is comparable to $l$. Accordingly, the correction changes the extremal and maximum-temperature scales and the local thermodynamic-stability structure; no independent dynamical evaporation endpoint is inferred from this plot alone.

\quad $\ast$ {\bf\textit{First Law and Horizon Entropy}}

To determine the Bekenstein-Hawking entropy\footnote{We adopt the Bekenstein--Hawking entropy law as a working assumption in our effective four-dimensional geometric description. Since the starting static metric incorporates effective LQC/braneworld corrections in the fluid sector of the Gao \textit{et al.} (2018) \cite{CHANGJUN} model, a rigorous entropy derivation from the Wald formalism would require the complete fundamental four-dimensional effective action. Higher-curvature terms could then generate additional quantum corrections to the area law. In the present work we retain $S=A_H/4$ as the usual low-energy effective prescription.} of the rotating metric, we calculate the area of the outer horizon by integrating the surface element on a constant-time section at $x=x_+$.
Using the Boyer--Lindquist form
\begin{eqnarray*}
	ds^2 = g_{tt} dt^2 + 2 g_{t\phi} dt d\phi + g_{xx} dx^2 + g_{\theta\theta} d\theta^2 + g_{\phi\phi} d\phi^2 ,
\end{eqnarray*}
and restricting to $dt=0$ and $dx=0$ on the horizon, the induced two-dimensional line element becomes
\begin{eqnarray*}
	ds_H^2 = g_{\theta\theta}(x_+, \theta) \, d\theta^2 + g_{\phi\phi}(x_+, \theta) \, d\phi^2 .
\end{eqnarray*}
For the REM,
\begin{eqnarray}
	g_{\theta\theta}(x_+, \theta) = \Sigma_+ = x_+^2 + a^2 \cos^2\theta
\end{eqnarray}
and
\begin{eqnarray}
	g_{\phi\phi}(x_+, \theta) = \frac{(x_+^2 + a^2)^2 \sin^2\theta}{\Sigma_+}.
\end{eqnarray}
The determinant of the induced horizon metric $\gamma_{\mu\nu}$ is
\begin{eqnarray}
	\det(\gamma) = g_{\theta\theta} \cdot g_{\phi\phi},
\end{eqnarray}
so the area element simplifies to
\begin{eqnarray}
	d{\bf A} = \sqrt{\det(\gamma)} \, d\theta \, d\phi
	= (x_+^2 + a^2) \sin\theta \, d\theta \, d\phi .
\end{eqnarray}
Integration over $\theta\in[0,\pi]$ and $\phi\in[0,2\pi]$ gives
\begin{eqnarray}
	{\bf A} = 4\pi (x_+^2 + a^2),
\end{eqnarray}
and Eq.~(\ref{T-S}) therefore yields
\begin{eqnarray}
	S = \frac{\bf A}{4} = \pi (x_+^2 + a^2).
	\label{Entropia-modificada}
\end{eqnarray}
The entropy written in terms of $x_+$ and $a$ has the same formal structure as the Bekenstein--Hawking entropy of a classical Kerr black hole. Nevertheless, the effective LQC/braneworld correction enters implicitly through the numerical value of the horizon radius $x_+$. 
Since $x_+$ is determined by the modified equation $\Delta(x_+)=0$, both the scale $l$ and the matter-sector index $f$ modify the horizon radius relative to the classical Kerr value $x_{+,\rm Kerr}=M+\sqrt{M^2-a^2}$. For $f=1$, this modifies the total area and entropy at fixed physical mass and angular momentum; for $f=2$ and $f=4$, the comparison is instead understood in terms of the corresponding effective state-space parameters.

\begin{figure}[htbp]
	\caption{Comparison between the classical and modified horizon entropies.}
	\centering
	\includegraphics[width=0.9 \textwidth, height=9cm, keepaspectratio]{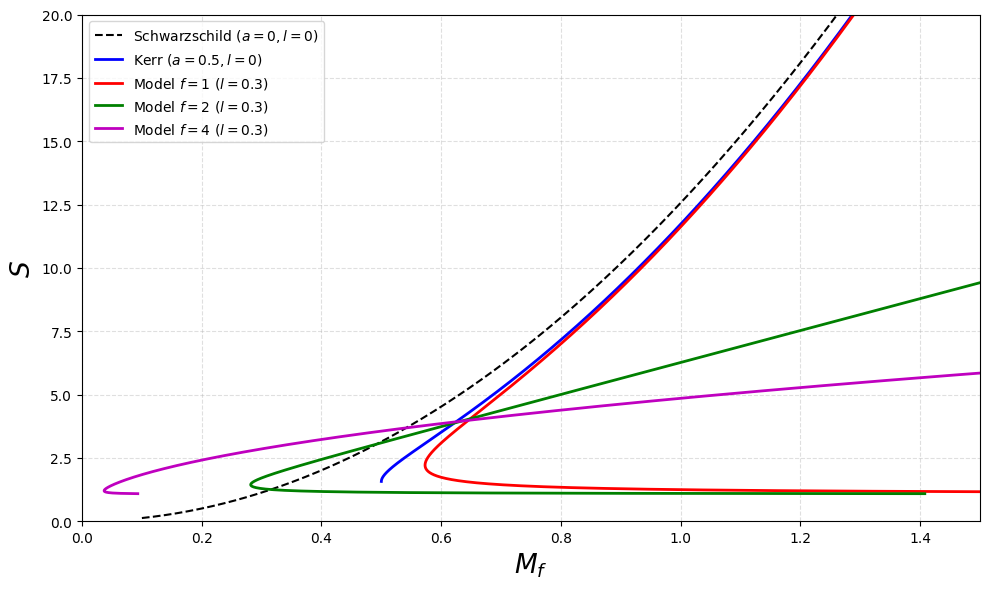}
	\label{entropia-1}
\end{figure}

For a general diffeomorphism-invariant theory, the entropy of a stationary horizon is given by the Wald prescription
\begin{eqnarray}
	S_{\text{Wald}} = -2\pi \int_{\mathcal{H}} \frac{\partial \mathcal{L}}{\partial R_{\alpha\beta\gamma\delta}} \epsilon_{\alpha\beta} \epsilon_{\gamma\delta} \sqrt{h} \, d\theta d\phi .
	\label{eq.wald-1}
\end{eqnarray}
Here $\epsilon_{\alpha\beta}=\kappa_H^{-1}\nabla_\alpha\chi_\beta$ is the normalized binormal on the horizon generated by $\chi^\mu=\partial_t+\Omega_H\partial_\phi$. For the purpose of the present effective treatment, we assume an Einstein--Hilbert gravitational sector, $\mathcal L=R/(16\pi)$, minimally coupled to the effective matter source. Under this explicit assumption,
\begin{eqnarray}
	\frac{\partial \mathcal{L}}{\partial R_{\alpha\beta\gamma\delta}}
	\epsilon_{\alpha\beta}\epsilon_{\gamma\delta}
	= -\frac{1}{8\pi},
	\label{eq.wald-2}
\end{eqnarray}
and the area law $S=A_H/4=\pi(x_+^2+a^2)$ is recovered. This is not a derivation of the fundamental LQC/braneworld action: if the complete four-dimensional effective Lagrangian contained higher-curvature or nonminimal curvature couplings, the Wald entropy would in general contain explicit additional corrections.
Figure~\ref{entropia-1} compares the classical and modified entropy curves; the reference plot uses $a=0.5$.
\FloatBarrier

The comparison reveals the combined influence of rotation, the effective correction scale, and the matter sector on the horizon state space. In the $f=1$ sector, where $M_1$ has the standard asymptotic mass interpretation, rotation suppresses the horizon area at fixed physical mass relative to Schwarzschild and therefore lowers the entropy. For $f=2$ and $f=4$, $M_f$ should instead be interpreted as an effective parameter of the radial matter profile, so that the corresponding curves provide a comparison in the effective parameter space rather than at a common physical ADM mass.

The scale $l$ further modifies the effective radial mass profile. For a given horizon radius $x_+$, a finite $l$ changes the mass $M_f$ required to sustain the same horizon area. Consequently, at fixed $M_f$, the modified branches can display a smaller effective horizon entropy than Kerr over the parameter range represented in the figure.

Finally, the response depends on the equation of state of the matter driving the collapse, parametrized by $f$. In the displayed comparison, the deviations from the classical Kerr curve become more pronounced as one moves from $f=1$ (dust) to $f=2$ (radiation) and $f=4$ (stiff matter). This behavior reflects the different radial matter profiles and their different contributions to the effective geometry.

\quad $\ast$ {\bf\textit{Extended First Law}}

To analyze the thermodynamic consistency of the rotating metric, we regard the horizon parameter $M_f(S,\mathcal{J}_f,l)$ as a state function on the extended state space $(S,\mathcal{J}_f,l)$, where $\mathcal{J}_f\equiv aM_f$. For the dust sector ($f=1$), $M_1$ admits the standard asymptotic mass interpretation and $\mathcal{J}_1=J$ is the physical angular momentum. For the $f=2$ and $f=4$ sectors, $M_f$ and $\mathcal{J}_f$ are instead regarded as effective state-space parameters. The exact differential is\footnote{In the original Gao \textit{et al.} (2018) \cite{CHANGJUN} dynamical-collapse parametrization, $l^{f+2}=3M_f/(4\pi\rho_{\rm cr})$ for fixed $\rho_{\rm cr}$. In the extended thermodynamic description used here, $l$ is promoted to an independent effective coupling. This is an extension of the original collapse parametrization, introduced so that variations of the effective high-density scale can be represented consistently in the state space $(S,\mathcal{J}_f,l)$.}
\begin{equation}
	dM_f = \left(\frac{\partial M_f}{\partial S}\right)_{\mathcal{J}_f,l} dS
	+ \left(\frac{\partial M_f}{\partial \mathcal{J}_f}\right)_{S,l} d\mathcal{J}_f
	+ \left(\frac{\partial M_f}{\partial l}\right)_{S,\mathcal{J}_f} dl .
	\label{eq:1a_lei_exata}
\end{equation}
Defining the thermodynamic conjugates
\begin{equation}
	T_{th} \equiv \left(\frac{\partial M_f}{\partial S}\right)_{\mathcal{J}_f,l},
	\quad
	\Omega_{th} \equiv \left(\frac{\partial M_f}{\partial \mathcal{J}_f}\right)_{S,l},
	\quad
	\Psi_l \equiv \left(\frac{\partial M_f}{\partial l}\right)_{S,\mathcal{J}_f},
	\label{f.estado-1}
\end{equation}
the corresponding extended state-space identity takes the exact form
\begin{equation}
	dM_f = T_{th} dS + \Omega_{th} d\mathcal{J}_f + \Psi_l dl .
\end{equation}
For $f=1$, where $\mathcal{J}_1=J$, this relation has the usual interpretation in terms of the physical angular momentum. For $f=2$ and $f=4$, it should instead be understood as an effective state-space identity, since $\mathcal{J}_f$ is not identified with an ADM angular momentum.

The conjugate potential associated with $l$ is
\begin{equation}
	\Psi_l = M_l \left( \frac{x_+ g}{x_+ g - a^2 g'} \right),
	\label{p.quantico-1}
\end{equation}
where $M_l\equiv(\partial M_f/\partial l)_{x_+,a}$ and
\begin{eqnarray}
	g(x_+) = \frac{x_+^{f-2}}{2\left(1 - \frac{l^{f+2}}{x_+^{f+2}}\right)}
	= \frac{x_+^{2f}}{2(x_+^{f+2} - l^{f+2})}.
	\label{g-1}
\end{eqnarray}
Because of the radial mass profile $g(x_+)$ and the finite correction scale $l$, the thermodynamic conjugates $T_{th}$ and $\Omega_{th}$ do not in general coincide identically with the geometric horizon quantities $T_H=\kappa_H/(2\pi)$ and $\Omega_H=a/(x_+^2+a^2)$. In the dust sector ($f=1$), this comparison has the standard thermodynamic interpretation because $\mathcal{J}_1=J$. For $f=2$ and $f=4$, $\Omega_{th}$ is instead the thermodynamic conjugate of the effective rotational state-space parameter $\mathcal{J}_f$.

In the dust sector, substituting the geometric relation $M_1=(x_+^2+a^2)g(x_+)$ into an attempted simple Kerr-like law $dM_1=T_HdS+\Omega_HdJ$ leads to the corresponding mismatch condition. More generally, the same algebraic condition provides a diagnostic of the departure of the effective state-space parametrization from the Kerr structure:
\begin{equation}
	x_+ g'(x_+) + g(x_+)
	= \frac{x_+^{f-2}}{2\left(1 - \frac{l^{f+2}}{x_+^{f+2}}\right)}
	\left[
	(f-1) - \frac{(f+2)l^{f+2}}{x_+^{f+2} - l^{f+2}}
	\right].
\end{equation}
Only in the classical Kerr limit, $l=0$ and $f=1$, is the condition $x_+g'+g=0$ identically satisfied, with $T_{th}\to T_H$ and $\Omega_{th}\to\Omega_H$, thereby recovering the standard Kerr first law. For finite $l$ in the $f=1$ sector, the differences $T_{th}-T_H$ and $\Omega_{th}-\Omega_H$ quantify the departure of the effective thermodynamic parametrization from the purely geometric horizon quantities. For $f=2$ and $f=4$, the same comparison is understood at the level of the effective state-space parametrization rather than as a standard ADM thermodynamic first law.

% BEGIN BLUE REVISION: Reviewer 1, coauthor horizon thermodynamics
{
\quad $\ast$ {\bf\textit{Pressure--Volume Contributions and the Interpretation of Enthalpy}}
\label{subsec:rem_pressure_volume}

The extended phase-space description of black-hole thermodynamics commonly
identifies the cosmological constant with a thermodynamic pressure,
$P_{\Lambda}=-\Lambda/(8\pi)$. When this pressure is allowed to vary, the
mass of an asymptotically anti-de Sitter black hole is interpreted as the
enthalpy, $H=U+P_{\Lambda}V_{\rm th}$, and the first law of an uncharged
rotating solution takes the form
\begin{equation}
 dH=T\,dS+\Omega\,dJ+V_{\rm th}\,dP_{\Lambda},
 \label{eq:rem_ads_extended_law}
\end{equation}
where $V_{\rm th}$ is the thermodynamic volume conjugate to
$P_{\Lambda}$ \cite{kastor2009enthalpy,kubiznak2017chemistry}.
This interpretation depends on specifying an extended family of solutions
in which the pressure is an independent thermodynamic variable. It does not
follow solely from the existence of an event horizon or from the presence
of effective matter. In particular, the rotating exterior metric (REM)
studied here contains no cosmological-constant term. Its modification of
the Kerr geometry is controlled instead by the effective length scale
$l$. Thus, a cosmological $V_{\rm th}\,dP_{\Lambda}$ term, together with
a universal identification $M_f=H$, cannot be inferred from the present
solution. This does not exclude other physically motivated notions of
pressure; it requires distinguishing them from the cosmological pressure
used in Eq.~\eqref{eq:rem_ads_extended_law}.

A pressure--volume work contribution can also arise directly from the
Einstein equations evaluated at a horizon. In this ``horizon
thermodynamics'' approach, pressure is associated with the radial
component of the matter stress tensor, and the relevant energy is a
quasi-local horizon energy rather than, in general, the mass interpreted
as enthalpy in extended phase-space thermodynamics
\cite{padmanabhan2002horizons,hansen2017horizon}.
To make this distinction explicit for the present model, we consider the
nonrotating limit $a=0$ of the three matter sectors $f=1,2,4$. The
metric then reads
\begin{equation}
 ds^2=-F_f(x)\,dt^2+\frac{dx^2}{F_f(x)}+x^2d\Omega^2,
 \qquad
 F_f(x)=1-\frac{2M_f}{x^f}
 \left(1-\frac{l^{f+2}}{x^{f+2}}\right).
 \label{eq:rem_static_pv}
\end{equation}
In the effective four-dimensional Einstein--Hilbert description adopted
for the area law in this work, we write
$G^{\mu}{}_{\nu}=8\pi T^{\mu}{}_{\nu,\mathrm{eff}}$.
The radial pressure is consequently defined by
$P_r(x)\equiv T^{x}{}_{x,\mathrm{eff}}$.
For Eq.~\eqref{eq:rem_static_pv}, the radial field equation gives
\begin{align}
 8\pi P_r(x)
 &=G^{x}{}_{x}
   =\frac{xF_f'(x)+F_f(x)-1}{x^2},
   \label{eq:rem_radial_einstein}\\
 P_r(x)
 &=\frac{M_f}{4\pi x^{f+2}}
   \left[(f-1)-(2f+1)
   \left(\frac{l}{x}\right)^{f+2}\right].
   \label{eq:rem_radial_pressure}
\end{align}
Equation~\eqref{eq:rem_radial_pressure} is a direct result of the
specified effective geometry; it does not introduce a new matter model
or an independently adjustable cosmological pressure. In particular,
for the dust sector $f=1$ it reduces to
\begin{equation}
 P_r(x)=-\frac{3M_1l^3}{4\pi x^6},
 \label{eq:rem_radial_pressure_dust}
\end{equation}
which vanishes in the Kerr/Schwarzschild limit $l\to0$. For $f=2$ and
$f=4$, the effective source generally remains nonzero even when $l=0$,
as expected from their distinct radial profiles.

Let $x_+$ be a positive, nondegenerate outer horizon of the static
metric, so that $F_f(x_+)=0$. Introducing the dimensionless ratio
\begin{equation}
 q_+\equiv\left(\frac{l}{x_+}\right)^{f+2},
 \qquad
 M_f=\frac{x_+^f}{2(1-q_+)},
 \label{eq:rem_horizon_q_pv}
\end{equation}
we obtain directly from the surface gravity and from
Eq.~\eqref{eq:rem_radial_pressure}
\begin{align}
 T_H&=\frac{F_f'(x_+)}{4\pi}
 =\frac{f-2(f+1)q_+}{4\pi x_+(1-q_+)},
 \label{eq:rem_static_temperature_pv}\\
 P_r(x_+)&=\frac{(f-1)-(2f+1)q_+}
 {8\pi x_+^2(1-q_+)}.
 \label{eq:rem_horizon_pressure_pv}
\end{align}
We adopt the area-law entropy used above, together with the geometric
horizon volume and the Misner--Sharp energy:
\begin{equation}
 S_H=\pi x_+^2,\qquad
 V_H=\frac{4\pi}{3}x_+^3,\qquad
 E_H=\frac{x_+}{2}.
 \label{eq:rem_horizon_ESV_pv}
\end{equation}
Combining Eqs.~\eqref{eq:rem_static_temperature_pv}--
\eqref{eq:rem_horizon_ESV_pv}, one finds, without further assumptions,
\begin{align}
 T_H\frac{dS_H}{dx_+}
 -P_r(x_+)\frac{dV_H}{dx_+}
 &=\frac{f-2(f+1)q_+}{2(1-q_+)}
  -\frac{(f-1)-(2f+1)q_+}{2(1-q_+)}
 \notag\\
 &=\frac12=\frac{dE_H}{dx_+}.
 \label{eq:rem_pressure_identity_check}
\end{align}
Hence the radial Einstein equation is exactly equivalent to the
horizon-displacement identity
\begin{equation}
 \boxed{dE_H=T_H\,dS_H-P_r\,dV_H},
 \label{eq:rem_horizon_pressure_first_law}
\end{equation}
for each of the three static sectors $f=1,2,4$.
Here $d$ denotes an infinitesimal displacement along the static
horizon family, and $P_r$ is evaluated at $x_+$. This is the
horizon-thermodynamic identity familiar from
Refs.~\cite{padmanabhan2002horizons,hansen2017horizon}, now made
explicit for the metric under consideration. It should not be
identified with the AdS enthalpy first law,
Eq.~\eqref{eq:rem_ads_extended_law}: the work term in
Eq.~\eqref{eq:rem_horizon_pressure_first_law} is
$-P_r\,dV_H$, not $V_{\rm th}\,dP_{\Lambda}$.

The distinction between quasi-local energy and the parameter $M_f$ is
also relevant. In the dust sector, where $M_1$ admits the conventional
asymptotic mass interpretation, the horizon equation gives
\begin{equation}
 M_1=\frac{x_+}{2\left(1-l^3/x_+^3\right)},
 \qquad E_H=\frac{x_+}{2}.
 \label{eq:rem_mass_vs_energy_pv}
\end{equation}
Thus $M_1\ne E_H$ for finite $l$. Indeed, differentiating the
asymptotic mass with respect to the area-law entropy at fixed $l$
yields
\begin{equation}
 \left(\frac{\partial M_1}{\partial S_H}\right)_l
 =\frac{T_H}{1-l^3/x_+^3},
 \label{eq:rem_mass_temperature_mismatch_pv}
\end{equation}
which agrees with the distinction between the state-space conjugate
$T_{\rm th}$ and the geometric Hawking temperature $T_H$ discussed
above. Such distinctions are particularly important when the effective
matter source depends explicitly on the mass parameter
\cite{ma2014regularfirstlaw,bokulic2021nonlinear}.
For $f=2$ and $f=4$, $M_f$ is not identified with an asymptotic ADM
mass in the present model, so an interpretation of $M_f$ as physical
enthalpy would require an additional definition of the corresponding
conserved energy.

Finally, the relation derived above is exact for $a=0$, but does not
by itself establish a pressure--volume first law for the rotating REM.
For $a\ne0$ the effective stress tensor is generally anisotropic and
angle dependent, and a rotating thermodynamic volume cannot be assumed
to equal the spherical expression in
Eq.~\eqref{eq:rem_horizon_ESV_pv}. Although horizon identities have
also been investigated for stationary axisymmetric geometries
\cite{kothawala2007thermodynamic}, a complete rotating construction
for the present metric would require identifying the appropriate
quasi-local or conserved energy and all independent work terms from a
specified effective action. Moreover, the original collapse
parametrization relates the two parameters through
\begin{equation}
 l^{f+2}=\frac{3M_f}{4\pi\rho_{\rm cr}},
 \label{eq:rem_density_constraint_pv}
\end{equation}
when the critical density $\rho_{\rm cr}$ is fixed. Treating $l$ as
independent, as in the preceding state-space discussion, is an
additional effective extension rather than a variation within the
original fixed-$\rho_{\rm cr}$ family.
For these reasons, we retain the exact parameter-space identity
$dM_f=T_{\rm th}\,dS+\Omega_{\rm th}\,d\mathcal{J}_f+\Psi_l\,dl$
without reinterpreting $M_f$ universally as enthalpy. The static
calculation demonstrates that a well-defined pressure--volume
contribution exists within the assumed effective Einstein description,
while indicating precisely what remains to be established for the
rotating solution.

}
% END BLUE REVISION

\quad $\ast$ {\bf\textit{Second Law}}

The second law of black hole mechanics is classically expressed by Hawking's area theorem, according to which the event-horizon area cannot decrease in physically admissible classical processes under the assumptions of the theorem, $\Delta A_H\ge0$. With the effective Bekenstein--Hawking identification $S=A_H/4=\pi(x_+^2+a^2)$, this corresponds to a nondecreasing horizon entropy, $\Delta S\ge0$. In the present rotating solution, $x_+$ is determined by $\Delta(x_+)=0$, whose relation to $M_f$, $a$, and $l$ depends explicitly on the matter sector $f$.

During processes such as matter absorption or rotational spin-down, the entropy variation is tied to the structure of the extended state-space relation and to the radial mass profile $g(x_+)$. For $f=1$, where the conventional asymptotic identifications $M_{\rm ADM}=M_1$ and $J=aM_1=\mathcal{J}_1$ apply, the horizon-area evolution is compatible with $\Delta S\ge0$ for physically admissible processes satisfying the appropriate energy conditions and the constraints of the corresponding thermodynamic relation. Finite-$l$ corrections continuously shift the corresponding horizon and entropy scales. We do not claim a general proof of the area theorem for arbitrary finite-$l$ processes.

For $f=2$ and $f=4$, additional conceptual care is required because $M_2$ and $M_4$ are not the standard ADM mass parameters. We therefore interpret them as effective parameters characterizing the radial matter profile rather than as conventional asymptotic masses. A full second-law analysis in these sectors would require a corresponding treatment of the admissible fluxes and energy conditions.

\subsubsection{Specific Heat}

%%%%%%%%%%%%%%%%%
{ 
The specific heat measures the local thermal response in a chosen
thermodynamic parametrization. For the dust sector ($f=1$), where
$M_1$ has the standard asymptotic mass interpretation, the natural
rotational variable is the physical angular momentum $J=aM_1$.
For the $f=2$ and $f=4$ sectors, however, $M_f$ is an effective
parameter rather than an ADM mass. In order to treat the three sectors
uniformly, we use the rotational state-space parameter introduced above,
\[
    \mathcal{J}_f \equiv aM_f .
\]
Thus, $\mathcal{J}_1=J$ for $f=1$, whereas for $f=2$ and $f=4$, 
$\mathcal{J}_f$ is understood as an effective state-space parameter,
without the standard asymptotic interpretation of physical angular
momentum.

We define the corresponding thermal-response function
$C_{\mathcal{J}_f}$. Its sign provides a diagnostic of the local
thermal response \cite{DAVIES,HOSEONG}:

$\circ$ $C_{\mathcal{J}_f}>0$: positive-heat-capacity branch.\\

$\circ$ $C_{\mathcal{J}_f}<0$: negative-heat-capacity branch.\\

$\circ$ $C_{\mathcal{J}_f}\to\infty$: Davies-type singular point.\\

The thermal-response function at fixed $\mathcal{J}_f$ is defined by
\begin{eqnarray}
    C_{\mathcal{J}_f}
    = T_H
    \left(
    \frac{\partial S}{\partial T_H}
    \right)_{\mathcal{J}_f}
    =
    T_H
    \frac{
    \left(
    \frac{\partial S}{\partial x_+}
    \right)_{\mathcal{J}_f}
    }{
    \left(
    \frac{\partial T_H}{\partial x_+}
    \right)_{\mathcal{J}_f}
    }.
    \label{Calor-1}
\end{eqnarray}
Since $\mathcal{J}_f=M_fa$ is held fixed during the differentiation,
$a$ varies along the constant-$\mathcal{J}_f$ curve. Therefore,}

%%%%%%%%%%%%%%%%%

%The specific heat of a rotating black hole measures its local thermal response in a chosen ensemble. Here we focus on the specific heat at fixed angular momentum, $C_J$. Its sign provides a diagnostic of local thermodynamic stability \cite{DAVIES,HOSEONG}:

%$\circ$ $C_J>0$: locally thermodynamically stable branch.\\

%$\circ$ $C_J<0$: locally thermodynamically unstable branch.\\

%$\circ$ $C_J\to\infty$: Davies-type critical point.\\
%The specific heat at fixed angular momentum is defined by
%\begin{eqnarray}
%	C_J = T_H \left( \frac{\partial S}{\partial T_H} \right)_J = T_H \frac{\left( \frac{\partial S}{\partial x_+} \right)_J}{\left( \frac{\partial T_H}{\partial x_+} \right)_J}.
%	\label{Calor-1}
%\end{eqnarray}
%Because $J=M_fa$ is held fixed during the differentiation, $a$ varies along the constant-$J$ curve. Therefore
%%%%%%%%%%%%%%%%%%
%%%%%%%%%%%%%%%%%%
%%%%%%%%%%%%%%%%%%
\begin{eqnarray}
	\left( \frac{\partial S}{\partial x_+} \right)_{\mathcal{J}_f} = 2\pi x_+ + 2\pi a \left( \frac{da}{dx_+} \right)_{\mathcal{J}_f} = 2\pi \left[ x_+ - \frac{a^2 M_x}{M_f + a M_a} \right]
	\label{Calor-2}
\end{eqnarray}
and, writing $T_x\equiv(\partial T_H/\partial x_+)_a$ and $T_a\equiv(\partial T_H/\partial a)_{x_+}$,
\begin{eqnarray}
	\left( \frac{\partial T_H}{\partial x_+} \right)_{\mathcal{J}_f} = T_x + T_a \left( \frac{da}{dx_+} \right)_{\mathcal{J}_f} = T_x - T_a \frac{a M_x}{M_f + a M_a},
\end{eqnarray}
where $g(x_+)$ is given by Eq.~(\ref{g-1}),
\begin{eqnarray}
	M_a = 2a g(x_+),
	\label{ma}
\end{eqnarray}
\begin{eqnarray}
	M_x = 2x_+ g(x_+) + (x_+^2 + a^2) g'(x_+)
	\label{mr}
\end{eqnarray}
and
\begin{eqnarray}
	M_f + a M_a = (x_+^2 + a^2) g + 2a^2 g = g(x_+) (x_+^2 + 3a^2).
	\label{mf}
\end{eqnarray}
Thus, Eq.~(\ref{Calor-1}) becomes
\begin{eqnarray}
	C_{\mathcal{J}_f} = T_H \frac{\left( \frac{\partial S}{\partial x_+} \right)_{\mathcal{J}_f}}{\left( \frac{\partial T_H}{\partial x_+} \right)_{\mathcal{J}_f}} = T_H \frac{2\pi \left[ x_+ - \frac{a^2 M_x}{M_f + a M_a} \right]}{T_x - T_a \frac{a M_x}{M_f + a M_a}}.
	\label{Calor-3}
\end{eqnarray}
{For the dust sector ($f=1$), one has $\mathcal{J}_1=J$, so that
$C_{\mathcal{J}_1}=C_J$ is the usual specific heat at fixed physical
angular momentum. For $f=2$ and $f=4$, $C_{\mathcal{J}_f}$ should be
understood as an effective thermal-response function at fixed
$\mathcal{J}_f=M_f a$.}
The Kerr result is recovered for $l\to0$ and $f=1$.

%%%%%%%%%%%%%%%%%%%%%%%
{The fixed-$\mathcal{J}_f$ thermal-response function reveals the local
thermal structure associated with the chosen state-space parametrization.
At an extremal point, where $T_H=0$, and assuming a regular denominator,
$C_{\mathcal{J}_f}$ vanishes. A divergence of
$C_{\mathcal{J}_f}$ occurs where the constant-$\mathcal{J}_f$
derivative of the temperature vanishes, defining a Davies-type
singular point.

For the Kerr limit ($f=1$, $l\to0$), one has
$\mathcal{J}_1=J$, and the usual fixed-angular-momentum
interpretation is recovered. For the reference value $a=0.5$, the
Davies point occurs at
$x_{\rm Davies}\simeq1.272$. A finite $l$ shifts its location.
The value $a=0.5$ labels the reference point/section used to evaluate
the resulting state-space function; the derivative defining
$C_{\mathcal{J}_f}$ itself is taken at fixed $\mathcal{J}_f$,
not at fixed $a$.

The singular point separates two local thermal-response branches:

\quad $\circ$ \textit{Positive-heat-capacity branch}
($x_+<x_{\rm Davies}$):

Between the extremal radius and the Davies point, the temperature
increases along the fixed-$\mathcal{J}_f$ branch and
$C_{\mathcal{J}_f}>0$.

\quad $\circ$ \textit{Negative-heat-capacity branch}
($x_+>x_{\rm Davies}$):

For radii above the Davies point, $C_{\mathcal{J}_f}<0$. In the
$f=1$ sector this reproduces the familiar negative-heat-capacity
behavior of asymptotically flat rotating black holes, whereas for
$f=2$ and $f=4$ it characterizes the corresponding effective thermal
response.

For $f=1$, where $\mathcal{J}_1=J$, the Davies point marks a change
in the local fixed-angular-momentum thermal response. For $f=2$ and
$f=4$, the corresponding singularities should instead be understood
as Davies-type features of the effective fixed-$\mathcal{J}_f$
parametrization. In either case, their interpretation as genuine
thermodynamic phase transitions would require a complementary analysis
of the appropriate thermodynamic potential and stability criteria.

Figure~\ref{calor-1} compares the modified and classical thermal-response
functions.

\begin{figure}[htbp]
    \caption{Comparison between the classical and modified thermal-response
    functions. For $f=1$, $C_{\mathcal{J}_1}=C_J$ is the usual
    fixed-angular-momentum specific heat, whereas for $f=2$ and $f=4$
    the curves represent the effective response at fixed
    $\mathcal{J}_f=M_fa$.}
    \centering
    \includegraphics[width=0.9\textwidth, height=9cm, keepaspectratio]{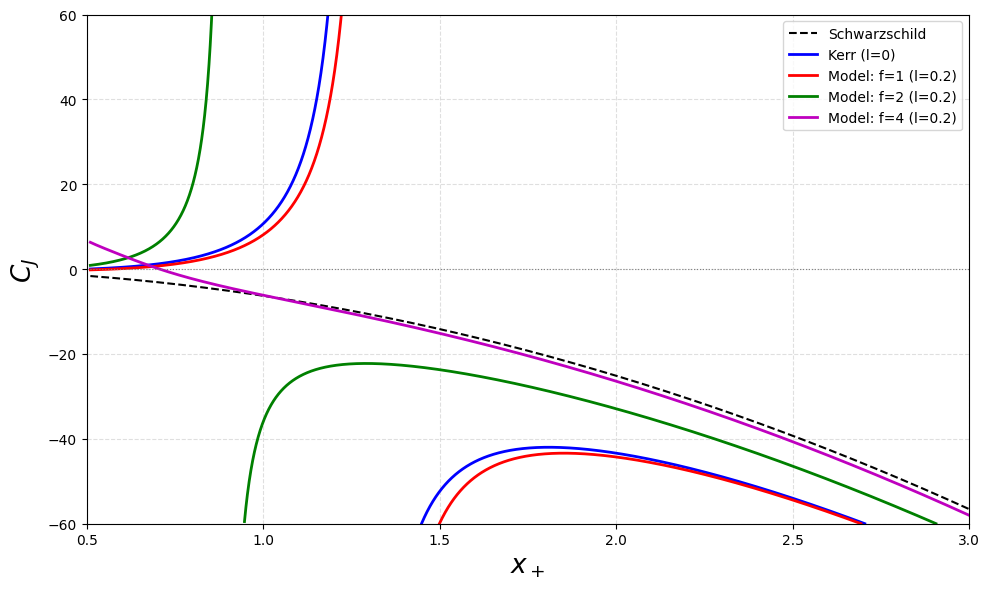}
    \label{calor-1}
\end{figure}
\FloatBarrier

For the parameter choices shown, a finite $l$ shifts the growth and
divergence structure of the thermal response relative to Kerr.}

%%%%%%%%%%%%%%%%%%%%%%%

%The fixed-$J$ specific heat reveals the local thermal stability structure. At an extremal point with $T_H=0$, and assuming a regular denominator, $C_J$ vanishes. The divergence of $C_J$ occurs where the constant-$J$ derivative of the temperature vanishes and defines a Davies-type critical point.

%For Kerr with the reference value $a=0.5$, this point occurs at $x_{\rm Davies}\simeq1.272$. Finite $l$ shifts its location. The value $a=0.5$ labels the reference point/section used to evaluate the resulting state-space function; the derivative defining $C_J$ itself is taken at fixed $J$, not at fixed $a$.

%The singular point separates two local branches:

%\quad $\circ$ \textit{Locally stable branch} ($x_+<x_{\rm Davies}$):

%Between the extremal radius and the Davies point, the temperature increases along the fixed-$J$ branch and $C_J>0$.

%\quad $\circ$ \textit{Locally unstable branch} ($x_+>x_{\rm Davies}$):

%For radii above the Davies point, $C_J<0$ and the branch displays the familiar negative-heat-capacity behavior of asymptotically flat black holes.

%The Davies point therefore marks a change in local thermodynamic stability. Its interpretation as a genuine phase transition requires a complementary analysis of the appropriate thermodynamic potential in the chosen ensemble.
%Figure~\ref{calor-1} compares the modified and classical specific heats.

%\FloatBarrier
%For the parameter choices shown, a finite $l$ shifts the growth and divergence structure of $C_J$ relative to Kerr.

%%%%%%%%%%%%%%%%%%%%%%%%%%%%%%%%%%%

\subsection{Causal Structure of the $f=2$ and $f=4$ Sectors and the Wormhole Hypothesis}

Interpreting the $f=2$ (radiation) and $f=4$ (stiff-matter) sectors as wormholes would require a separate global and geometric analysis. A traversable wormhole requires, in particular, a throat --- a minimum-area two-surface connecting two regions --- satisfying the flare-out condition \cite{Morris1988,Visser1995,Lobo2008}. In classical GR, maintaining such a throat generally requires violation of the null energy condition \cite{Morris1988,Hochberg1997}. In the present rotating model, the horizon candidates are determined by $\Delta(x_\pm)=0$.

For the parameter ranges in which Eq.~(\ref{II-2}) admits two positive real roots, $x_+$ and $x_-$ define outer and inner Killing horizons, respectively, producing a causal structure formally reminiscent of Reissner--Nordström \cite{CHANGJUN}. A complete identification of a global event horizon would require the corresponding global causal extension.

Within the diagnostics considered here, the $f=2$ and $f=4$ solutions do not display the defining features of traversable wormholes:

\begin{itemize}
	\item[$\circ$] \textbf{No open flare-out throat:} the area radius does not exhibit a strict minimum with $R''(x_0)>0$ in a horizon-free region.
	\item[$\circ$] \textbf{Killing horizons:} the surfaces $x_\pm$ indicate an effective rotating black hole-like geometry rather than a traversable bridge between two asymptotic regions.
	\item[$\circ$] \textbf{Asymptotic falloff:} for $f>1$, $g_{tt}+1$ falls faster than $1/x$, so $M_2$ and $M_4$ should be regarded as effective parameters of the radial matter profile rather than as the standard ADM mass.
\end{itemize}

Accordingly, we do not identify the $f=2$ and $f=4$ sectors as wormholes. We refer to them conservatively as effective rotating black hole-like geometries with inner and outer Killing horizons, subject to the qualifications above.

%%%%%%%%%%%%%%%%%%%%%%  Final Remarks  %%%%%%%%%%%%%%%%%%%%%%%%%%%%%

\section{Final Remarks}

General Relativity provides a remarkably successful relativistic description of gravitation, but the singular behavior of classical collapse solutions motivates the study of effective high-density corrections. The Schwarzschild solution and its rotating generalizations remain the essential benchmarks against which such models should be tested.

Gao \textit{et al.} (2018) extended the Oppenheimer--Snyder collapse picture by incorporating effective LQC/braneworld corrections. In their model the density can reach a critical scale, halting collapse and allowing a comoving expansion phase, thereby generating a nonsingular pulsating interior dynamics.

We have applied the modified Newman--Janis construction in the Azreg-Aïnou formulation to the GLSF exterior metric. The resulting REM is a stationary rotating exterior geometry associated with that pulsating-collapse model. Its horizon equation, angular velocity, surface gravity, and Hawking temperature consistently recover the Kerr and Schwarzschild limits when $l\to0$.

The horizon entropy has been treated under the explicit effective assumption that the four-dimensional exterior gravitational sector is Einstein--Hilbert with minimally coupled effective matter, so that the Wald prescription reduces to $S=A_H/4$. The fixed-$\mathcal{J}_f$ thermal-response function exhibits Davies-type singular points separating positive- and negative-heat-capacity branches. For $f=1$, $\mathcal{J}_1=J$ and this reduces to the usual fixed-angular-momentum specific heat $C_J$.

The simple Kerr-like first law does not hold generically at finite $l$. In the extended state space, with $l$ promoted to an independent effective coupling, the exact state-function identity is $dM_f=T_{\rm th}dS+\Omega_{\rm th}d\mathcal{J}_f+\Psi_l\,dl$. The $f=1$ sector admits the conventional asymptotic interpretation of $M_1$ and $\mathcal{J}_1=J=aM_1$ and  its area behavior is compatible with the second law for physically admissible processes under the appropriate assumptions. For $f=2$ and $f=4$, $M_f$ and $\mathcal{J}_f$ are instead regarded as effective state-space parameters; these sectors require an effective matter-profile interpretation and are not identified here as wormholes.

%%%%%%%%%%%%%%%%%%

\acknowledgments We thank CNPq and CAPES for partial financial support. F.A.B. acknowledges support from CNPq (Grant no. $309092/2022-1$).

%%%%%%%%%%%%%%%%%%%%%%  References %%%%%%%%%%%%%%%%%%%%%%%%%%%%%%%%%%%%%%%%%%%

\appendix

\section{Horizon Entropy and the Effective Einstein--Hilbert Assumption}

The entropy of a stationary horizon in a diffeomorphism-invariant gravitational theory can be computed using the Noether--Wald prescription \cite{WALD}:
\begin{equation}
	S_{\text{Wald}} = -2\pi \int_{\mathcal{H}} \frac{\partial \mathcal{L}}{\partial R_{\alpha\beta\gamma\delta}} \epsilon_{\alpha\beta} \epsilon_{\gamma\delta} \sqrt{h}\,d\theta\,d\phi
\end{equation}
where $\epsilon_{\alpha\beta}=\kappa_H^{-1}\nabla_\alpha\chi_\beta$ denotes the normalized binormal on the horizon $\mathcal H$, generated by $\chi^\mu=\partial_t+\Omega_H\partial_\phi$. To evaluate the entropy one must specify the effective four-dimensional gravitational action. For the purposes of the present effective treatment, we \emph{assume} that the exterior geometry can be represented by an Einstein--Hilbert gravitational sector minimally coupled to an effective matter source:
\begin{equation}
	\mathcal{S}_{\text{eff}} = \int d^4x \sqrt{-g} \left[ \mathcal{L}_{\text{grav}} + \mathcal{L}_{\text{matter}}(g_{\mu\nu}, \Psi) \right]
\end{equation}
where $\mathcal L_{\rm grav}=R/(16\pi)$ and $\mathcal L_{\rm matter}$ represents the effective matter sector associated with the modified radial mass profile $M(x)$. Under the explicit minimal-coupling assumption, $\partial\mathcal L_{\rm matter}/\partial R_{\alpha\beta\gamma\delta}=0$, and the functional derivative with respect to the Riemann tensor comes entirely from the Einstein--Hilbert term:
\begin{equation}
	\frac{\partial \mathcal{L}}{\partial R_{\alpha\beta\gamma\delta}} = \frac{1}{32\pi} \left( g^{\alpha\gamma} g^{\beta\delta} - g^{\alpha\delta} g^{\beta\gamma} \right)
\end{equation} Contracting with the binormal, which satisfies $\epsilon_{\alpha\beta}\epsilon^{\alpha\beta}=-2$, gives
\begin{equation}
	\frac{\partial \mathcal{L}}{\partial R_{\alpha\beta\gamma\delta}} \epsilon_{\alpha\beta} \epsilon_{\gamma\delta} = -\frac{1}{8\pi}
\end{equation} Substitution into the Wald formula yields the Bekenstein--Hawking area law:
\begin{equation}
	S_{\text{Wald}} = \frac{1}{4} \int_{\mathcal{H}} \sqrt{h}\,d\theta\,d\phi = \frac{A_H}{4} = \pi(x_+^2 + a^2)
\end{equation} Thus, within this effective Einstein--Hilbert/minimal-coupling assumption, $S_{\rm Wald}=A_H/4$. This should not be interpreted as a derivation of the fundamental LQC/braneworld gravitational action. If the fundamental effective action contains higher-curvature terms or nonminimal curvature couplings, the Wald entropy would in general receive additional corrections.

%%%%%%%%%%%%%%%%%%%%%%  References %%%%%%%%%%%%%%%%%%%%%%%%%%%%%%%%%%%%%%%%%%%

\end{document}